\documentclass[
 reprint, superscriptaddress,
 amsmath,amssymb,
 aps,
]{revtex4-2}
\usepackage{graphicx}
\usepackage{dcolumn}
\usepackage{bm}
\usepackage{amsmath}
\usepackage[normalem]{ulem}
\usepackage{xcolor}
\begin{document}
\preprint{APS/123-QED}
\title{Near-field enhancement by a metasurface at octupole plasmon resonance in periodic disc dimers}

\author{Sagar Sehrawat}
\email{sagar.sehrawat@aalto.fi}
\affiliation{Department of Applied Physics, Aalto University, P.O. Box 13500, FI-00076 Aalto, Finland}

\author{Klas Lindfors}
\affiliation{Department of Applied Physics, Aalto University, P.O. Box 13500, FI-00076 Aalto, Finland}
\affiliation{Institute for Light and Matter, University of Cologne, Greinstrasse 4-6, 50939 Cologne, Germany}

\author{Andriy Shevchenko}
\affiliation{Department of Applied Physics, Aalto University, P.O. Box 13500, FI-00076 Aalto, Finland}

\begin{abstract}
Local intensity enhancement by plasmonic nanoparticles is widely used in optics and photonics. However, the effect is usually based on dipole resonances in the particles. Recently, it has been shown that quadrupole and octupole resonances can exhibit comparable, or even higher near-field enhancement. In this work, we focus on the near-field enhancement by a metasurface composed of gold-disc dimers arranged in a rectangular array. We find that, owing to an octupole plasmon resonance coupled to a surface lattice resonance, exceptionally high near-field enhancement in the dimer gaps can be achieved in the visible spectral range. To gain insight into the effect, we develop an analytical model for the effective dipole and octupole polarizabilities of the particles in an array, and discover, that at decreasing array periods, the dipole polarizability tends to vanish, while the octupole polarizability rapidly increases. Hence, octupole resonances can find applications in high-density arrays of plasmonic resonators. We propose a method to numerically evaluate multipole polarizabilities of a single particle, applying it to the gold dimer that we consider. The influence of the array on the effective polarizabilities is then verified by numerical calculations and a good agreement is obtained. Our results may open new avenues for investigating the properties of periodic plasmonic structures based on higher-order multipole resonances and their applications.
\end{abstract}

\maketitle

\section{Introduction}
Plasmonic nanoparticles have been extensively studied and widely employed in optics and photonics because of their unique optical properties, including near-field enhancement and efficient scattering of optical fields. Central to these properties are the localized surface plasmon resonances (LSPRs), which are collective oscillations of conduction electrons at the surface of a metal particle excited by an optical field \cite{maier2007plasmonics,amendola2017surface}. The resonances are highly dependent on the material composition of the particles, which makes them tunable in a wide spectral range. Being sensitive to small variations in the refractive index of the surrounding medium, LSPRs provide an excellent platform for applications in optical sensing, imaging, and spectroscopy \cite{hutter2004exploitation,haes2005detection,nie1997probing,endo2005label,raschke2003biomolecular,haes2004unified,petryayeva2011localized}. When nanoparticles form a periodic lattice, the fields scattered by them can interfere constructively at their own locations, leading to another type of resonances called surface lattice resonances (SLRs) \cite{carron1986resonances,kravets2018plasmonic,zou2005silver,bin2021ultra,miroshnichenko2010fano,le2019enhanced}. SLRs are often narrowband and have a high quality factor. For metal particles, SLRs can overlap with LSPRs of individual nanoparticles, leading to an additional near-field enhancement of light intensity \cite{kolkowski2023enabling,zang2023near,rodriguez2013surface,sadeghi2016tunable,shi2022significant,hooper2018second}. Due to these properties, plasmonic nanoparticle arrays have been successfully used, e.g., in biochemical sensing \cite{rodriguez2013surface,sadeghi2016tunable,thackray2014narrow}, nonlinear optics \cite{huttunen2018using,michaeli2017nonlinear,czaplicki2018less}, laser technology \cite{hakala2017lasing,wang2020lasing}, and spectroscopy \cite{adato2009ultra,oh2016engineering,theiss2010plasmonic,kravets2018plasmonic}. In general, the higher is the density of the particles on the surface, the stronger should be the overall optical signal received from them. However, in plasmonic metasurfaces with sub-wavelength periods, SLRs cannot be excited (since for them, the lattice period should be comparable with or larger than the wavelength used, $\lambda$), while LSPRs weaken when the separation approaches 0, since the particles start to behave as a continuous film. This means that the density of the particles is limited to $ca.$ $\lambda^{-2}$ in a two-dimensional array.
\par Both LSPRs and SLRs are usually treated in terms of electric dipole excitations in the particles, and higher-order multipole resonances are rarely considered, mainly when studying bound states in the continuum and other dark excitations \cite{habteyes2011theta,bauer2015towards}. Indeed, for subwavelength particles, higher-order multipoles are nearly absent in the far-field compared to electric dipoles. However, it has been recently shown that, in individual scatterers, higher-order multipoles, such as quadrupoles and octupoles, can lead to an exceptionally high near-field enhancement \cite{sehrawat2024octupole,sehrawat2024near}. In addition, these multipole excitations can hybridize with the electric dipole excitation and result in both narrowband and bright far-field scattering \cite{sehrawat2024hybridization}. These findings show promise for many practical applications and also open up prospects for the discovery of new phenomena concerning higher-order multipole excitations in arrayed particles.

\par In this work, we study scattering and near-field enhancement of light by a two-dimensional array of metal particles and find that the array periods can be tuned to significantly enhance the hybrid dipole-octupole LSPR through its coupling to an SLR. Choosing the particles in the form of gold-disc dimers, we show that the coupling leads to an exceptionally high intensity enhancement in the dimer gaps. The enhancement is tuned to appear near $\lambda$ = 700 nm, making it suitable for biomedical sensing applications (using SERS and other plasmon-enhanced spectroscopic methods) and detectable using standard silicon detectors. Moreover, we have discovered, that while the dipole polarizability vanishes when the array period decreases below the wavelength, the octupole polarizability rapidly grows, allowing for a high density of hot spots in the array. The effect is revealed in our analytical calculations of the effective polarizabilities of the particles. We propose a method to numerically calculate the dipole, quadrupole, and octupole polarizabilities of individual particles and apply it to verify our analytical calculations. Our results underline the importance of higher-order multipole resonances in arrayed optical scatterers, demonstrating their ability to efficiently enhance and scatter optical fields in the visible spectral range. In our studies, we used the classical and scattering-current multipole expansion \cite{grahn2012electromagnetic}, implementing them with the help of COMSOL Multiphysics. 
\section{Scattering-current multipole expansion and multipole polarizabilities}
The classical multipole expansion is an important technique employed in electromagnetic theory to describe the fields scattered by small particles, as these fields can be decomposed into orthogonal components produced by point multipoles located at the center of the scatterer. However, the expansion does not provide a clear picture of the excitations in the particle. Some of them, such as toroidal multipole and anapole excitations, are not revealed by the expansion. To expose the actual excitations in the scatterer, one can expand the scattering current density 
\begin{equation} \label{Js}
    \textbf{J}(\textbf{r})=-i\omega \epsilon_0 \bigr[\epsilon(\textbf{r})-\epsilon_{\text{s}}\bigr] \textbf{E}(\textbf{r})
\end{equation}
into orthogonal Cartesian current multipoles \cite{grahn2012electromagnetic}. Here, $\epsilon(\textbf{r})$ is the relative electric permittivity whose value outside the scatterer is $\epsilon_\text{s}$. Note that the scattering current density includes also the displacement current density and is not equal to zero for dielectric scatterers. The current multipoles are not divided into electric and magnetic multipoles and are represented by very simple current configurations \cite{grahn2012electromagnetic}. The expansion includes toroidal multipoles as multipoles of higher orders. For example, the toroidal electric dipole is in the current expansion a combination of octupoles. The expansion is applicable to arrayed scatterers, revealing the dependence of the multipole excitations in them on the neighboring scatterers. In terms of point current multipoles, the localized scattering current density can be written as \cite{grahn2012electromagnetic}
\begin{widetext}
\begin{equation} \label{J}
    \textbf{\text{J}}(\textbf{\text{r}})=i\omega \sum_{l=1}^\infty \sum_{\hat{\textbf{\textit{v}}}=\hat{\textbf{\textit{x}}}, \hat{\textbf{\textit{y}}}, \hat{\textbf{\textit{z}}}} \sum_{a=0}^{l-1} \sum_{b=0}^{l-a-1} M^{(l)}(\hat{\textbf{\textit{v}}},a,b)\hat{\textbf{\textit{v}}} \frac{(-1)^l (l-1)!}{a!b![l-(a+b+1)]!} \frac{\text{d}^a}{\text{d}x^a}\frac{\text{d}^b}{\text{d}y^b}\frac{\text{d}^{l-(a+b+1)}}{\text{d}z^{l-(a+b+1)}}\delta(\textbf{r}),
\end{equation}
\end{widetext}

where $M^{(l)}(\hat{\textbf{\textit{v}}},a,b)$ is the current multipole moment of order $l$, with $\hat{\textbf{\textit{v}}}$ being a unit vector along the direction of the current density, and $a$ and $b$ showing how many times the current density reverses its direction when crossing the particle in the $x$- and $y$-direction, respectively. Along the $z$-direction, the current density flips $l - (a + b + 1)$ times. For scatterers that are much smaller than the wavelength, the multipole moments can be calculated as
\begin{equation}
\begin{split}
M^{(l)}(\hat{\textbf{\textit{v}}},a,b)=\frac{i}{\omega(l-1)!}\int{\!J_v(\textbf{r})x^a y^b z^{(l-a-b-1)}\text{d}^3r},
\end{split}
\end{equation}
where $J_v(\textbf{r})$ is the $v$-component of the current density. Beyond the small-scatterer (or long-wavelength) approximation, the current multipole moments are calculated from the classical electric and magnetic multipole moments $a_\text{E}(l,m)$ and $a_\text{M}(l,m)$ using the mapping relations presented in Ref. \cite{grahn2012electromagnetic}. The classical multipole moments are in turn calculated from the actual scattering-current distribution in the particle obtained, e.g., numerically (see Eqs. (15) and (16) in \cite{grahn2012electromagnetic}, where the related numerical calculations are demonstrated as well). The effective amplitude of the current density associated with multipole moment $M^{(l)}(\hat{\textbf{\textit{v}}},a,b)$ scales as $-i\omega \hat{\textbf{\textit{v}}} M^{(l)}(\hat{\textbf{\textit{v}}},a,b)/(x_\text{e}^a y_\text{e}^b z_\text{e}^{l - a -b - 1})$, where $x_\text{e}$, $y_\text{e}$, and $z_\text{e}$ are the effective sizes of the scattering current distribution along the $x$-, $y$-, and $z$-direction \cite{sehrawat2024octupole, sehrawat2024near}. These effective sizes can be considered as fitting parameters when the contributions of individual multipole moments to the near-field enhancement spectra have to be evaluated \cite{sehrawat2024octupole, sehrawat2024near}. The near-field enhancement is proportional to the effective amplitude of the localized current density that enhances the field.
\begin{figure*}[ht]
    \centering
    \includegraphics[width=\textwidth]{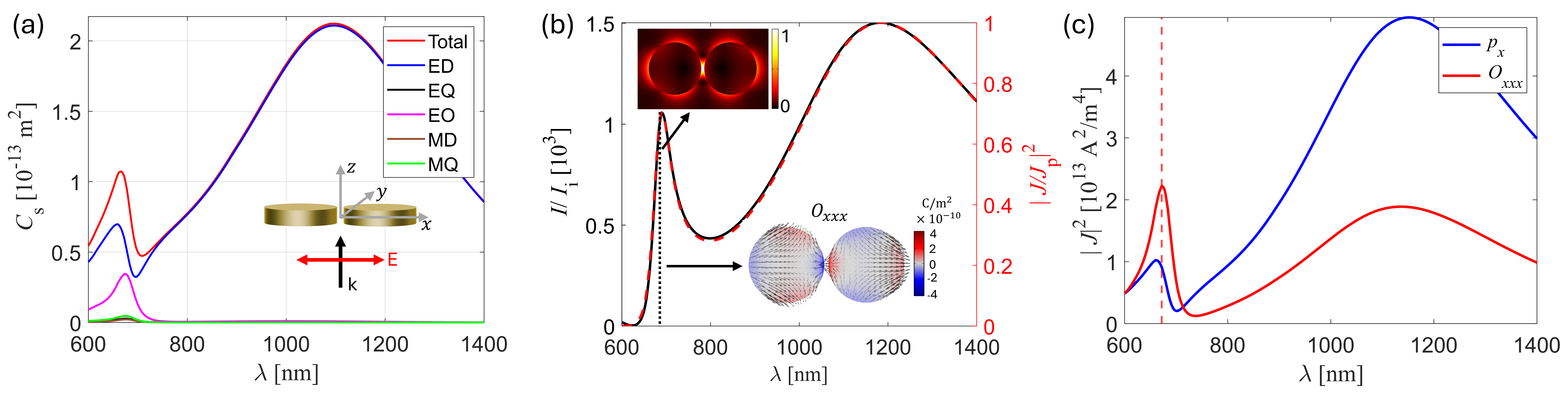}
    \caption{(a) The spectrum of the scattering cross section $C_s$ (red line) of a gold disc dimer shown in the inset. The contribution to $C_s$ of the electric dipole, electric quadrupole, electric octupole, magnetic dipole, and magnetic quadrupole are shown by the blue, black, magenta, brown, and green lines, respectively. (b) Intensity enhancement factor (solid black line) and scattering current density (red dashed line). (c) The squared absolute values of the dipole (blue line) and octupole (red line) current densities in the dimer. The red dashed line marks the wavelength of 672 nm.}
    \label{singledimer}
\end{figure*}
Once the current multipole moments are calculated, one can find the polarizabilities of the particle related to these moments. In general, the dipole polarizability is a second-rank tensor that can be made diagonal by rotating the coordinate system such that its axes become the symmetry axes of the particle. In this case, the three diagonal components of the dipole polarizability, $\alpha_v$, satisfy the expression
\begin{equation}
    p_v = M^{(1)}(\hat{\textbf{\textit{v}}},0,0) = \alpha_v E_v,
\end{equation}
where $p_v$ is the Cartesian component of the dipole moment. The quadrupole polarizability is a third-rank tensor with 27 elements. Choosing the coordinate system to be aligned with the symmetry axes of the particle and dealing with quadrupole moments $Q_{vx} = M^{(2)}(\hat{\textbf{\textit{v}}},1,0), Q_{vy} = M^{(2)}(\hat{\textbf{\textit{v}}},0,1)$, and $Q_{vz} = M^{(2)}(\hat{\textbf{\textit{v}}},0,0)$, one can reduce the number of non-zero elements of the tensor from 27 to 9 and write them as $\beta_{vi}$ \cite{grahn2013multipole}. However, quadrupoles are known to be sensitive to the field gradient rather than to the field itself, with each $Q_{vi}$ sensitive to $\text{d}E_v/\text{d}i$. Considering a finite-sized particle centered at the origin of the coordinate system, we can use a Taylor series expansion of the field with respect to coordinate $i$ about $i = 0$. Truncating the series after the second term we obtain $E_v(i) = E_v(0) + i \text{d}E_v(0)/\text{d}i$. Using Eqs. (1) and (3), one can find that the first term in the expression for $M^{(2)}(\hat{\textbf{\textit{v}}}, a, b)$ vanishes for centrosymmetric particles. When volume-averaged in Eq. (3), the expanded field can be written as $E_v(0) +\space i_\text{e}\text{d}E_v(0)/\text{d}i$, where $i_\text{e}$ plays the role of effective radial size of the scatterer along the $i$-direction. Since $i_\text{e}$ is a parameter to be determined, we introduce another parameter, $\beta^{\text{(i)}}_{vi} = \beta_{vi}i_\text{e}$, that contains $i_\text{e}$ and write
\begin{equation}
    Q_{vi} = \beta_{vi}E_v +\beta^{\text{(i)}}_{vi}\text{d}E_v/\text{d}i,
\end{equation}
where the argument of $E_v$ is dropped for brevity. Without the second term, quadrupole excitations by inhomogeneous fields would be ignored. The first term can include excitations of quadrupole moments by inhomogeneous intraparticle fields, e.g., of the excited dipole moment. Thus, two parameters, $\beta_{vi}$ and $\beta^{\text{(i)}}_{vi}$, can be considered to characterize quadrupole excitations in a given scatterer. A similar expression can be written for an octupole moment $O_{vij}$ when the coordinate system is aligned with respect to the symmetry axes of the particle:
\begin{equation}
    O_{vij} =\! \gamma_{vij}E_v +\! \gamma^{\text{(i)}}_{vij}\frac{\text{d}E_v}{\text{d}i} +\! \gamma^{\text{(ii)}}_{vij}\frac{\text{d}E_v}{\text{d}j} 
+\! \gamma^{\text{(iii)}}_{vij} \frac{\text{d}}{\text{d}i}\bigg(\frac{\text{d}E_v}{\text{d}j}\bigg).
\end{equation}
Formally, also terms $\gamma^{\text{(iv)}}_{vij}\text{d}^2E_v/\text{d}i^2 $ and $\gamma^{\text{(v)}}_{vij}\text{d}^2E_v/\text{d}j^2 $ must be present in the expression, but we neglect them for the case of $i \neq j$ as terms containing second spatial derivatives; if $i = j$, then these terms are automatically included in the fourth term, being represented by $\gamma^{\text{(iii)}}_{vij}$. The number of terms in the above expression can in some cases be further reduced by considering the actual geometry of the particle. For example, the terms $\gamma^{\text{(i)}}_{vij}\text{d}E_v/\text{d}i$ and $\gamma^{\text{(ii)}}_{vij}\text{d}E_v/\text{d}j$ vanish for centrosymmetric particles. In any case, one can see that, for evaluation of the polarizabilities, the particle must be considered separately in the absence and in the presence of spatial derivatives of the incident field. In section IV, we show how to evaluate the relevant dipole and octupole polarizabilities of metal disc dimers.
\section{Metal disc dimer}
\begin{figure*}[ht]
    \centering
    \includegraphics[width=\textwidth]{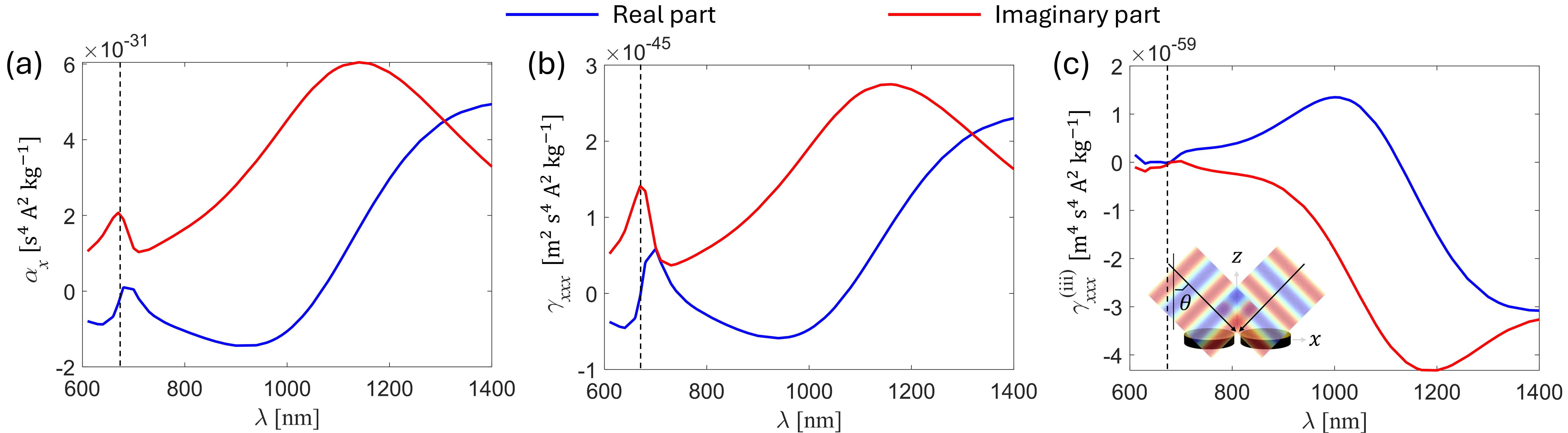}
    \caption{The spectra of the numerically calculated polarizabilities (a) $\alpha_x$, (b) $\gamma_{xxx}$, and (c) $\gamma^{\text{(iii)}}_{xxx}$ for the considered dimer. The blue and red lines represent the real and imaginary parts of the polarizabilties, respectively. The inset in (c) shows the illumination of the dimer by two interfering plane waves for calculating $\gamma^{\text{(iii)}}_{xxx}$. The vertical dashed black line marks the wavelength of 672 nm.}
    \label{NPol}
\end{figure*}
As a structure with a non-negligible octupole resonance in the visible spectral range, we consider a dimer consisting of two gold discs, each with a radius of 90 nm and a thickness of 40 nm, separated from each other by a gap of 10 nm. Metal dimers of this type are frequently used in surface enhanced Raman spectroscopy (SERS), with the highest local-field enhancement achieved in the gap between the discs \cite{zang2023near,qin2006designing,shevchenko2012large,rycenga2010understanding}. Here, we consider the dimer to be located in glass, which has a refractive index $n =$ 1.5. The optical constants of gold are taken from Ref. \cite{johnson1972optical}. In our simulations, we assume that the dimer is illuminated with a plane wave polarized along the longest dimension of the structure (the $x$-axis) and propagating along the shortest dimension (the $z$-axis), as shown in the inset of Fig. \ref{singledimer}(a). The geometrical parameters are selected such that the octupole resonance appears at the edge of the visible spectral range (meeting the requirements for applications in SERS and making the observation of the resonance possible with standard silicon detectors). Figure \ref{singledimer}(a) shows the spectra of the scattering cross-section of the dimer ($C_s$; red line) and the contributions of all the relevant electric and magnetic multipole moments to it. The spectra were calculated using Eqs. (15) and (16) of Ref. \cite{grahn2012electromagnetic}, and the required scattering current density in the dimer was calculated using COMSOL Multiphysics. The electric-dipole (ED; blue line) contribution prevails in the infrared range, while the contribution of the electric octupole (EO; magenta line) is noticeable at $\lambda <$ 700 nm. The contributions of the electric quadrupole (black line), magnetic dipole (brown line), and magnetic quadrupole (green line) are seen to be considerably weaker.

\par Since the near-field amplitude of the scattered light is proportional to the current density in the particle, we next calculate the current densities associated with the two most pronounced multipoles, which are the $x$-polarized dipole ($p_x$) and $x$-polarized octupole ($O_{xxx}$). Here, we deal with current multipole moments obtained from the classical multipole expansion coefficients using the following equations:
\begin{equation}
    p_x=-\frac{1}{C_1}\bigg(a_\text{E}(1,1)+\frac{7}{3}a_\text{E}(3,1)\bigg),
\end{equation}
\begin{equation}
    O_{xxx}=-\frac{1}{3C_3}a_\text{E}(3,1),
\end{equation}
where we have $C_1 = -ik^3/(6\pi\epsilon E_0)$ and $C_3 = -ik^5/(210\pi \epsilon E_0)$. These equations are the same as Eqs. (6) and (8) in Ref. \cite{sehrawat2024near}. The amplitudes of the current densities associated with these multipole moments can be written as
\begin{equation}
    \textbf{J}_{p_x}=-i\omega\hat{\textbf{x}}p_x s,
\end{equation}
\begin{equation}
    \textbf{J}_{O_{xxx}}=-2i\omega\hat{\textbf{x}}O_{xxx} \frac{s}{x_\text{e}^2},
\end{equation}
where $s$ is a proportionality coefficient and $x_\text{e}$ the effective radial extent of the current density in the $x$-direction (see Eqs. (2) and (4) in \cite{sehrawat2024near}). The sum of the current densities at the center of the dimer, where the currents of the dipole and octupole oscillate out of phase, is given by
\begin{equation}
    \textbf{J}=-i \omega s \hat{\textbf{x}}\Big(p_x-2 \frac{O_{xxx}}{x_\text{e}^2}\Big).
\end{equation}
The parameter $x_\text{e}$ is considered as a fitting parameter found by matching the spectrum of the magnitude of $ \textbf{J}$ to the spectrum of the field enhancement factor evaluated numerically. Figure \ref{singledimer}(b) shows a normalized intensity enhancement ($I/I_\text{i}$; solid black line) in the gap between the discs. It exhibits a broad dipole resonance centered at $\lambda = 1200$ nm and a weaker and narrowband hybrid resonance at 692 nm. The latter includes the dipole and octupole excitations. By adjusting $x_\text{e}$ we obtain the current density spectrum shown by the dashed red line in Fig. \ref{singledimer}(b). We used a linear dependence of $x_\text{e}$ on the wavelength with its values at the wavelengths of 692 and 1200 nm equal to 109 and 124 nm, respectively. The spectra of the individual current densities of dipole and octupole excitations are shown in Fig. \ref{singledimer}(c) by the blue and red lines, respectively. The vertical red dashed line at $\lambda =$ 672 nm marks the peak of the octupole current density that exceeds the dipole current density at this wavelength. At $\lambda \approx$ 1200 nm, however, the dipole current density is dominant.
\par The two of the relevant polarizabilities $\alpha_x$ and $\gamma_{xxx}$ are calculated using Eqs. (4) and (6), respectively, for the case of the considered plane-wave illumination [see the inset of Fig. \ref{singledimer}(a)]. Note that the derivatives of the incident plane wave with respect to $x$ are equal to 0, which removes the three last terms from Eq. (6). The spectra of $\alpha_x$ and $\gamma_{xxx}$ are shown in Figs. \ref{NPol}(a) and \ref{NPol}(b), respectively. In general, the excitation of the octupole moment $O_{xxx}$ in our centrosymmetric dimer is sensitive to the second derivative of the incident field, $\text{d}^2E_x/\text{d}x^2$, rather than to $\text{d}E_x/\text{d}x$. Hence, it can be approximated as $O_{xxx} = \gamma_{xxx}E_x + \gamma^\text{(iii)}_{xxx}\text{d}^2E_x/\text{d}x^2$. The polarizability $\gamma^\text{(iii)}_{xxx}$ can be calculated by making the dimer interact with a standing optical wave, e.g. formed by two plane waves as shown in the inset of Fig. \ref{NPol}(c). Evaluating $O_{xxx}$ (numerically with COMSOL Multiphysics) allows us to find $\gamma^\text{(iii)}_{xxx}$ from Eq. (6), since $\gamma_{xxx}$ is already known. In the calculations, we use the incident angle $\theta$ of the two plane waves equal to 10 deg and set their phases equal at the center of the dimer. Figure \ref{NPol}(c) shows the spectrum of $\gamma^\text{(iii)}_{xxx}$. The calculated polarizabilities are used later on in the paper, when considering an array of such gold dimers.
\section{Dimers in an array}
By arranging the dimers of section III in a two-dimensional periodic array, we aim to obtain additional near-field enhancement at the octupole resonance wavelength due to multiple scattering of light in the array. The enhancement can be described by the increase of the effective polarizabilities of the particles as the incident field is modified by the fields scattered by the other particles. We start by developing a simple analytical model for the effective polarizabilities (subsection A) and proceed to exact numerical studies of the dimers in an array (subsection B).
\subsection{Analytical model for effective dipole and octupole polarizabilities}
\begin{figure}[t]
    \centering
    \includegraphics[scale=0.86]{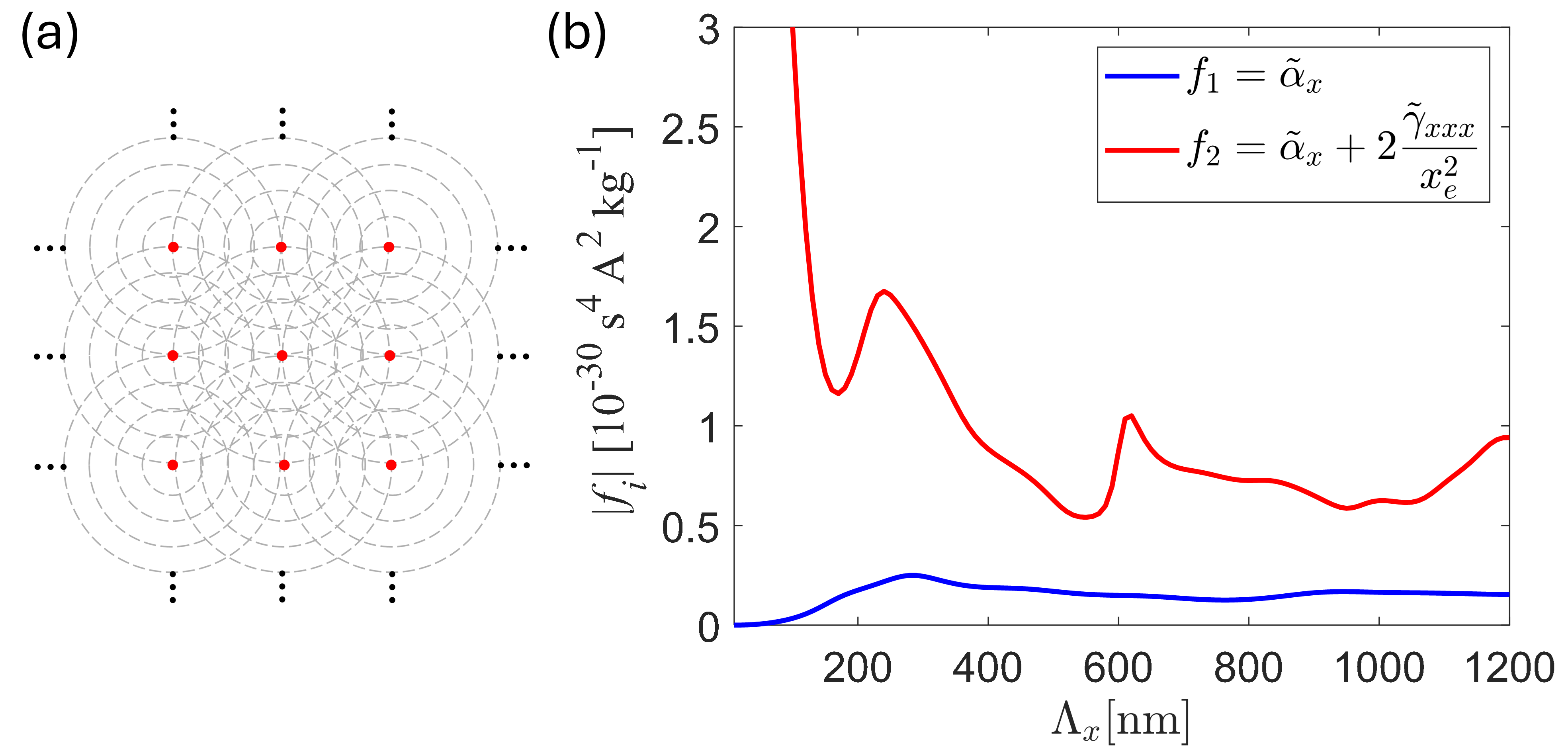}
    \caption{(a) Point scatterers (red dots) in an infinite periodic array. The dashed lines show schematically the fields scattered by the particles. (b) The magnitude of the effective dipole polarizability ($\tilde{\alpha}_x$; blue line) at $\lambda$ = 1184 nm and the sum $\tilde{\alpha}_x+2\tilde{\gamma}_{xxx}/x_\text{e}^2$ (red line) at $\lambda$ = 672 nm.}
    \label{toy}
\end{figure}

Consider a system of periodically arranged point-like scatterers with periods $\Lambda_x$ and $\Lambda_y$ in the $x$ and $y$ directions, respectively. The scatterers are presented by the red dots in Fig. \ref{toy}(a), and the circular dashed lines in the figure schematically show the spherical-like scattered waves of the particles. We assume that the array is illuminated by an $x$-polarized plane wave at normal incidence and that only dipole and octupole moments $p_x$ and $O_{xxx}$ are excited in the scatterers. By symmetry, the total scattered field at the position of any particle is $x$-polarized as well. The fields scattered by the particles via dipole and octupole excitations can be separated. At the position of any individual particle, the total complex amplitudes of these separated fields can be written in terms of the (actual) excited dipole and octupole moments as
\begin{equation}
    E_\text{d} = \tilde{p}_x\zeta_\text{d} = \tilde{\alpha}_x E_\text{i}\zeta_\text{d},
\end{equation}
\begin{equation}
    E_\text{o} = (\tilde{\gamma}_{xxx}E_\text{i} + \tilde{\gamma}^\text{(iii)}_{xxx}\text{d}^2E_\text{i}/\text{d}x^2)\zeta_\text{o},
\end{equation}
where $E_\text{i}$ is the incident field amplitude, the quantities with the tilde are the actual (or effective) quantities that take into account the presence of the fields scattered by the other particles. The functions $\zeta_\text{d}$ and $\zeta_\text{o}$ are the spatial distribution functions of the scattered fields taking into account the propagation directions and distances in the array (see the Appendix). They depend on $\Lambda_x$ and $\Lambda_y$, as will be shown below. The second term in Eq. (13) is equal to zero, since for the incident plane wave, we have $\text{d}^2E_\text{i}/\text{d}x^2$ = 0. The dipole moment excited in each particle is calculated as
\begin{equation}
\begin{split}
    \Tilde{p}_x & =\alpha_x (E_\text{i}+E_\text{d}+E_\text{o})\\
    &=\alpha_x E_\text{i} (1+\Tilde{\alpha}_x \zeta_\text{d} + \Tilde{\gamma}_{xxx} \zeta_\text{o})=\Tilde{\alpha}_x E_\text{i},
\end{split}
\end{equation}
which implies that $\Tilde{\alpha}_x = \alpha_x \left(1 + \Tilde{\alpha}_x \zeta_\text{d} + \Tilde{\gamma}_{xxx} \zeta_\text{o} \right)$. When calculating the effective dipole polarizability $\Tilde{\alpha}_x$, one can neglect the highly confined octupole fields (which are relatively dark in the far-field) and obtain
\begin{equation}
    \Tilde{\alpha}_x=\frac{\alpha_x}{1-\alpha_x \zeta_\text{d}} (1+\Tilde{\gamma}_{xxx} \zeta_\text{o})\approx\frac{\alpha_x}{1-\alpha_x \zeta_\text{d}}.
\end{equation}
The octupole moment $\Tilde{O}_{xxx}$ can be written in a similar way in terms of the effective polarizabilities:
\begin{equation} \label{tildaoxxx}
\begin{split}
     \Tilde{O}_{xxx} & =\gamma_{xxx} E_\text{i} \Big(1+\Tilde{\alpha}_x \zeta_\text{d}+\Tilde{\gamma}_{xxx} \zeta_\text{o} \Big)\\
     &+\gamma^\text{(iii)}_{xxx} E_\text{i} \Biggr(\Tilde{\alpha}_x \frac{\partial^2}{\partial x^2} \zeta_\text{d} +\Tilde{\gamma}^\text{(iii)}_{xxx} \frac{\partial^2}{\partial x^2} \zeta_\text{o}\Biggr)=\tilde{\gamma}_{xxx}E_\text{i}.
\end{split}
\end{equation}
Using this result and Eq. (15), we obtain
\begin{multline}\label{tilda_gamma_slash}
    \Tilde{\gamma}_{xxx} =
    \Big( \gamma_{xxx}+\gamma^\text{(iii)}_{xxx} \alpha_x \zeta_\text{d}^{''} \Big) \bigg/ \\
    \Big( (1-\gamma_{xxx} \zeta_\text{o}-\gamma^\text{(iii)}_{xxx} \zeta_\text{o}^{''})(1-\alpha_x \zeta_\text{d}) \\
    -\alpha_x \zeta_\text{o} (\gamma_{xxx} \zeta_\text{d}+\gamma^\text{(iii)}_{xxx} \zeta_\text{d}^{''}) \Big)
\end{multline}
Again, assuming the scattered octupole fields to be weak compared to the dipole fields, we set $\zeta_\text{o}$ and $\zeta''_\text{o}$ to 0 and obtain
\begin{equation}
    \Tilde{\gamma}_{xxx}\approx\frac{\gamma_{xxx} +\gamma^\text{(iii)}_{xxx} \alpha_x \zeta_\text{d}^{''}}{1-\alpha_x \zeta_\text{d}}.
\end{equation}
Hence, for the scattered dipole fields dominating over the octupole fields at the position of each scatterer, the evaluation of the effective polarizabilities $\alpha_x$ and $\gamma_{xxx}$ requires knowledge of the dipole-field distribution function $\zeta_\text{d}$ only. This function at a point shifted from the particle in question by a distance $x$ along the $x$-axis is (see the Appendix)
\begin{widetext}
    \begin{equation} \label{zd}
\begin{split}
     \zeta_\text{d} & =\sum_{n = -\frac{N - 1}{2},\space (n\neq0)}^{\frac{N - 1}{2}} \left(\frac{k^2}{2 \pi \epsilon_0 \epsilon |n\Lambda_x-x|}\left[\frac{1}{k^2 (n\Lambda_x-x)^2} - \frac{i}{k |n \Lambda_x-x|}\right] e^{i k |n\Lambda_x-x|}\right)+\\
     &\sum_{n=-\frac{N-1}{2}}^{\frac{N-1}{2}} \Bigg(\sum_{m=1}^{\frac{M-1}{2}}\Bigg(\frac{k^3}{2\pi \epsilon \epsilon_0}\Bigg[\frac{2(n\Lambda_x-x)^2+(m\Lambda_y)^2}{k^3 \Big((n\Lambda_x-x)^2+(m\Lambda_y)^2\Big)^{3/2}}-\frac{i(2(n\Lambda_x-x)^2+(m\Lambda_y)^2)}{k^2((n\Lambda_x-x)^2+(m\Lambda_y)^2)}\\
     &-\frac{(m\Lambda_y)^2}{k\sqrt{(n\Lambda_x-x)^2+(m\Lambda_y)^2}}\Bigg]\frac{1}{(n\Lambda_x-x)^2+(m\Lambda_y)^2}e^{ik\sqrt{(n\Lambda_x-x)^2+(m\Lambda_y)^2}}\Bigg)\Bigg).
\end{split}
\end{equation}
\end{widetext}
For a single dimer, the intensity enhancement spectrum has been shown to exhibit two peaks corresponding to the dipole and hybrid dipole-octupole plasmon resonances. At the wavelength of the dipole resonance (1184 nm), the absolute value of the dipole polarizability $\tilde{\alpha}_x$ is shown as a function of $\Lambda_x$ by the blue line in Fig. \ref{toy}(b) for the case of $\Lambda_y$ = 300 nm. In Eq. (19), $x$ was set to zero, and both $N$ and $M$ were set to 9, making the array to contain 81 scatterers. This size of the array is large enough to be equivalent to an infinite array when evaluating the effective polarizability of the particle at the center. The effective polarizability stays approximately constant at large values of $\Lambda_x$, but when $\Lambda_x$ decreases towards 0, it vanishes. This shows that the density of "hotspots" in a plasmonic array cannot be made arbitrarily high, because of their self-quenching. At the wavelength of the hybrid resonance (672 nm), we calculate the quantity $|\Tilde{\alpha}_x + 2\Tilde{\gamma}_{xxx}/x_\text{e}^2|$ that combines the contributions of both polarizabilities $\tilde{\alpha}_x$ and $\tilde{\gamma}_{xxx}$ to the overall current density in the particle. This quantity is represented by the red line in Fig. \ref{toy}(b) for the case of $x_\text{e}$ = 110 nm obtained previously for this wavelength. As the array period decreases, the total polarizability at the hybrid resonance grows, in contrast to the dipole polarizability. This suggests that plasmonic hotspots due to hybrid dipole-octupole excitations become even "hotter" when their density increases, which is a surprising and potentially useful discovery in view of many applications, including surface-enhanced fluorescence and Raman spectroscopy.
In the above calculations, absorption of the scattered spherical waves by the other particles was ignored. To take this into account, we introduce a complex wavevector for these waves, with the imaginary part of it determined from the expression
\begin{equation}
    \kappa_i=\frac{1}{2\Lambda_i}\text{ln}\Biggr(\frac{I_1}{I_2}\Biggr).
\end{equation}
Here, $i$ is equal to $x$ or $y$ for the wave propagation along the $x$- or $y$-axis, respectively. When evaluating $\kappa_i$ numerically, we consider a plane wave instead of a spherical wave and calculate the local intensity of the wave on the array surface before and after propagating over one unit cell, obtaining $I_1$ and $I_2$ for Eq. (20) (see the next subsection). Then, in Eq. (19) we replace $k$ with $2\pi/\lambda (n+ i\kappa)$, where $\kappa = (\kappa_x + \kappa_y)/2$ only in the arguments of the exponential functions.
\begin{figure}[b]
    \centering
    \includegraphics[width=0.45\textwidth]{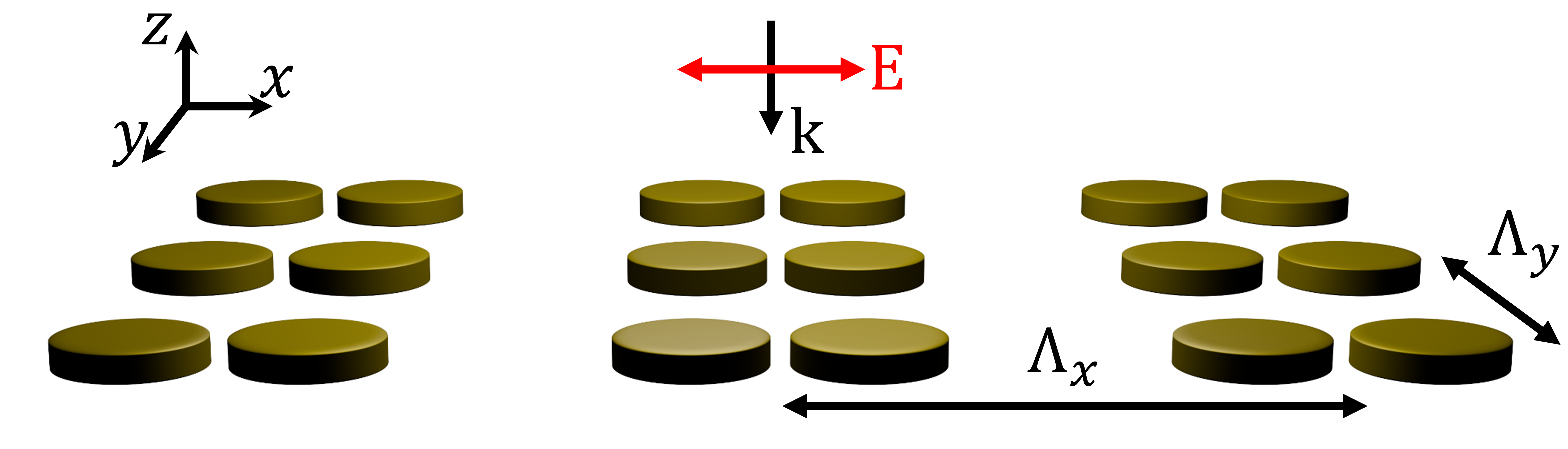}
    \caption{The interaction geometry of the considered gold-dimer metasurface with an incident $x$-polarized plane wave.}
    \label{geom}
\end{figure}
\begin{figure*}[t]
    \centering
    \includegraphics[width=\textwidth]{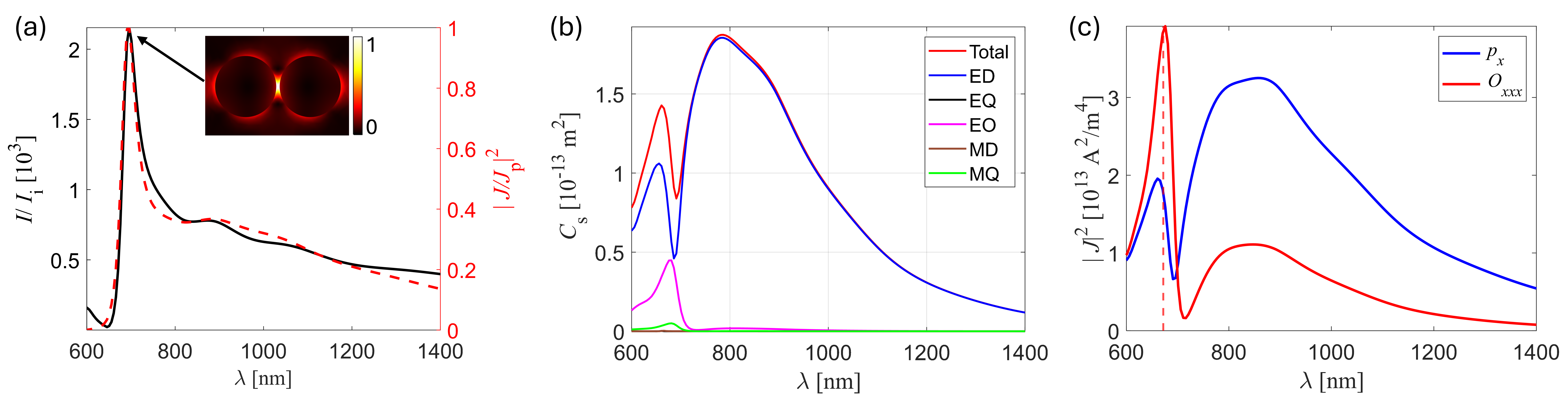}
    \caption{(a) The spectra of the near-field intensity enhancement factor (solid black line) and the squared absolute value of the normalized current density (dashed red line) calculated for a dimer in an optimized array. The inset shows the electric field amplitude distribution at the peak wavelength of 692 nm. (b) The total scattering cross section (red line) of a dimer in the array and the contributions to it from the constituent multipoles (blue, black, magenta, brown, and green lines corresponding to the electric dipole, electric quadrupole, electric octupole, magnetic dipole, and magnetic quadrupole). (c) The squared absolute value of the dipole (blue line) and octupole (red line) contributions to the scattering current density. The dashed line marks the wavelength of 672 nm.}
    \label{dimerarray}
\end{figure*}
\subsection{Numerical studies}
The array of gold dimers considered in our numerical studies is shown in Fig. \ref{geom}. It is embedded in glass and has periods $\Lambda_x$ and $\Lambda_y$ in the $x$ and $y$ directions, respectively. The incident wave propagates along the $z$-axis and is polarized along the $x$-direction. We optimize the periods $\Lambda_x$ and $\Lambda_y$ to be 430 nm and 300 nm, respectively, for the gap enhancement in the dimers to be highest at the wavelength of the octupole resonance. The intensity enhancement spectrum calculated for the optimized structure is shown in Fig. \ref{dimerarray}(a) by the solid black line. The scattering current density obtained with the effective radial extension $x_\text{e}$ is shown by the dashed red line. At the peak, we have $x_\text{e} =$ 109 nm. For dimers in the array, the hybrid dipole-octupole resonance is seen to result in a higher enhancement, such that the dipole resonance becomes nearly negligible. The changes are attributed to the collective optical response of the dimers to the incident wave. The normalized near-field amplitude distribution of one of the dimers is shown in the inset. In Fig. \ref{dimerarray}(b), we show the spectrum of the scattering cross-section of a dimer in the array (see the red line) and the contributions of the relevant multipoles to this spectrum. As before, the electric-dipole contribution (blue line) to this spectrum is dominating, and the electric-octupole contribution (magenta line) is almost equal to it at $\lambda =$ 690 nm. The dipole peak is seen to be blue-shifted compared to that of a single dimer. The squared absolute values of the current density associated with the dipole and octupole moments are plotted in Fig. \ref{dimerarray}(c). At the wavelength of 672 nm (marked with the vertical dashed red line), the octupole current density (solid red line) is twice as large as the dipole one (blue line). Indeed, when resonantly supported by the lattice response of the array, the octupole excitation stays relatively dark. However, the current density and the near-field enhancement associated with it significantly increase, which is not the case for the dipole resonance.
\begin{figure}[b]
    \centering
    \includegraphics[scale=0.50]{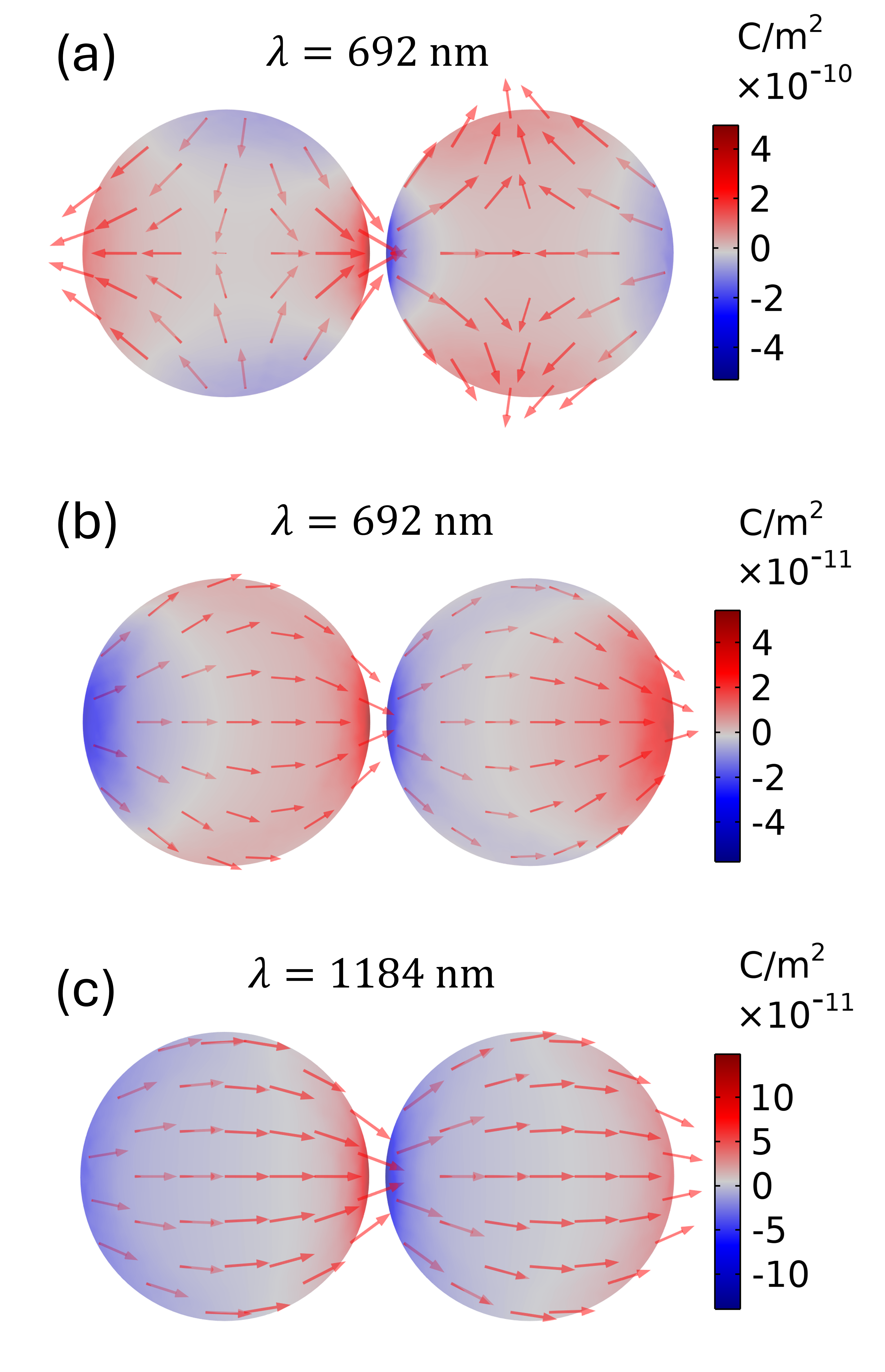}
    \caption{The distribution of the charge and current densities on the surface of a disc dimer in an array for the two characteristic wavelengths of 692 nm in (a) and (b), and 1184 in (c). The distribution in (a) shows the peak current density of the octupole excitation, while in (b), a dipole excitation shifted in phase by $\pi/2$ is observed. In (c), corresponding to $\lambda =$ 1184 nm, we find a dipole excitation only.}
    \label{profiles}
\end{figure}
\par At the wavelengths of 692 nm and 1184 nm, the current and surface charge densities exhibit the distributions shown in Fig. \ref{profiles}. In (a), corresponding to $\lambda =$ 692 nm, a clear octupole excitation is observed, including the moments $O_{xxx}$ and $O_{yxy}$. Note that the latter does not contribute to the gap enhancement. At the same wavelength, the dipole excitation coexists with the octupole one, being phase-shifted by $\pi/2$ shown in Fig. \ref{profiles}(b). The octupole charge density is seen to be an order-of-magnitude higher than the dipole charge density. In Fig. \ref{profiles}(c), corresponding to $\lambda =$ 1184 nm, only the current dipole can be observed.
\begin{figure}[b]
    \centering
    \includegraphics[scale=0.5]{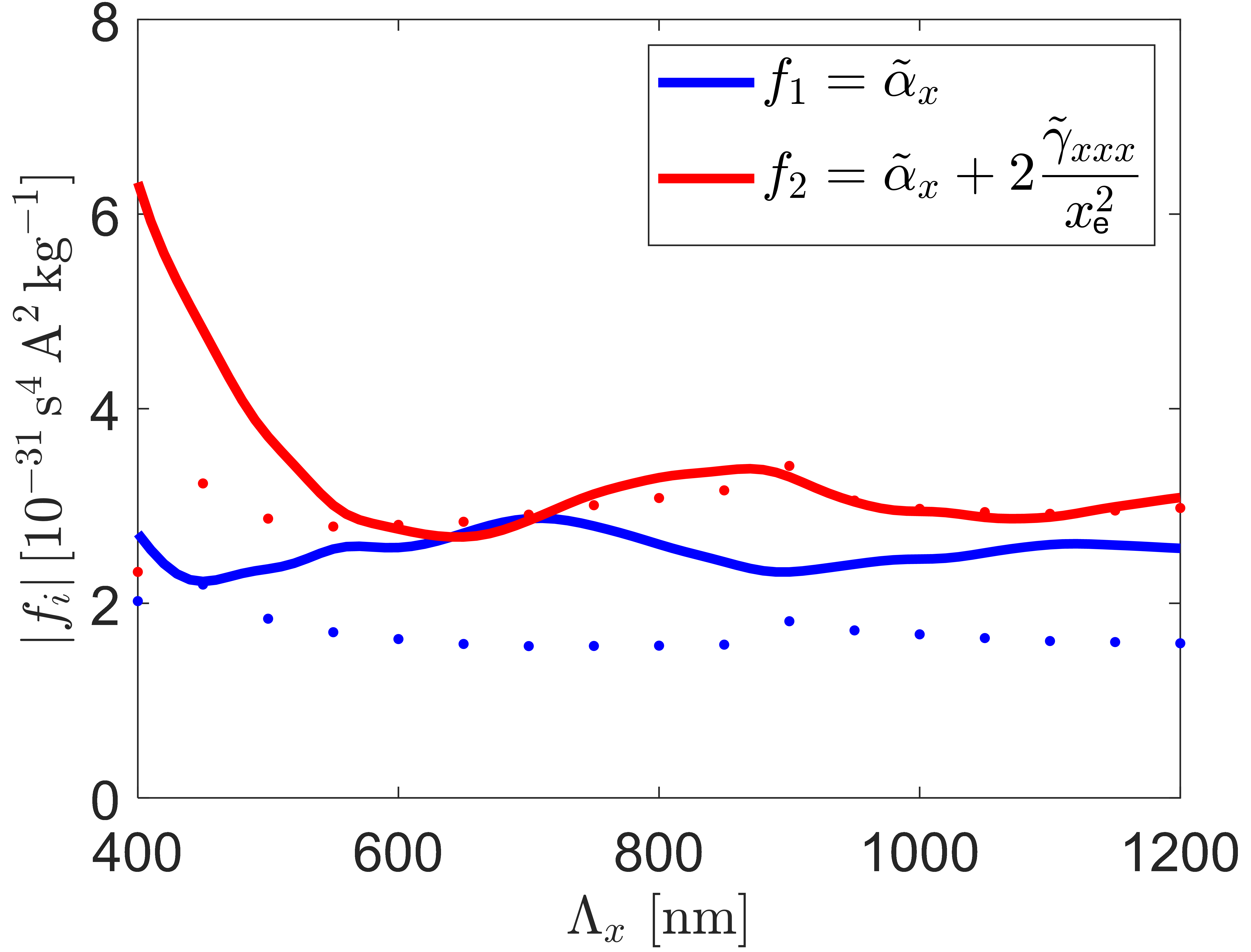}
    \caption{The effective dipole polarizability $|\Tilde{\alpha}_x|$ (blue solid and dotted lines) and properly scaled octupole polarizability $|2 \Tilde{\gamma}_{xxx}/x_\text{e}^2|$ (red solid and dotted lines). The solid and dotted lines correspond to the analytical and numerical calculations.}
    \label{analytic}
\end{figure}
\par Next, we calculate the effective dipole and octupole polarizabilities, $\tilde{\alpha}_x$ and $\tilde{\gamma}_{xxx}$, for the arrayed dimers both numerically (from the values of $p_x$ and $O_{xxx}$) and analytically with the help of Eq. (\ref{zd}) averaged over the total size of the dimer in the $x$-direction. The period $\Lambda_y$ is fixed at 300 nm, while the period $\Lambda_x$ is varied. In Fig. \ref{analytic}, we show the dipole (blue) and scaled octupole (red) polarizabilities as functions of $\Lambda_x$ obtained numerically (dots) and analytically (solid lines) at $\lambda =$ 692 nm. The scaled octupole polarizability, $2\tilde{\gamma}_{xxx}/x_\text{e}^2$, can be compared with $\tilde{\alpha}_x$, because the associated current densities are proportional to these quantities. The dipole polarizability stays below the scaled octupole polarizability for all values of $\Lambda_x$. Additionally, the octupole polarizability can be seen to increase when $\Lambda_x$ decreases toward zero. This observation supports our finding that octupole excitations become stronger at high densities of the particles in the array. At $\Lambda_x =$ 400 nm, the numerically evaluated octupole polarizability becomes low compared to its analytically calculated value because the period starts to approach the size of the dimer. While the two calculation approaches do not yield identical results, they show a reasonable agreement.
\begin{figure}[b]
    \centering
    \includegraphics[width=0.75\linewidth]{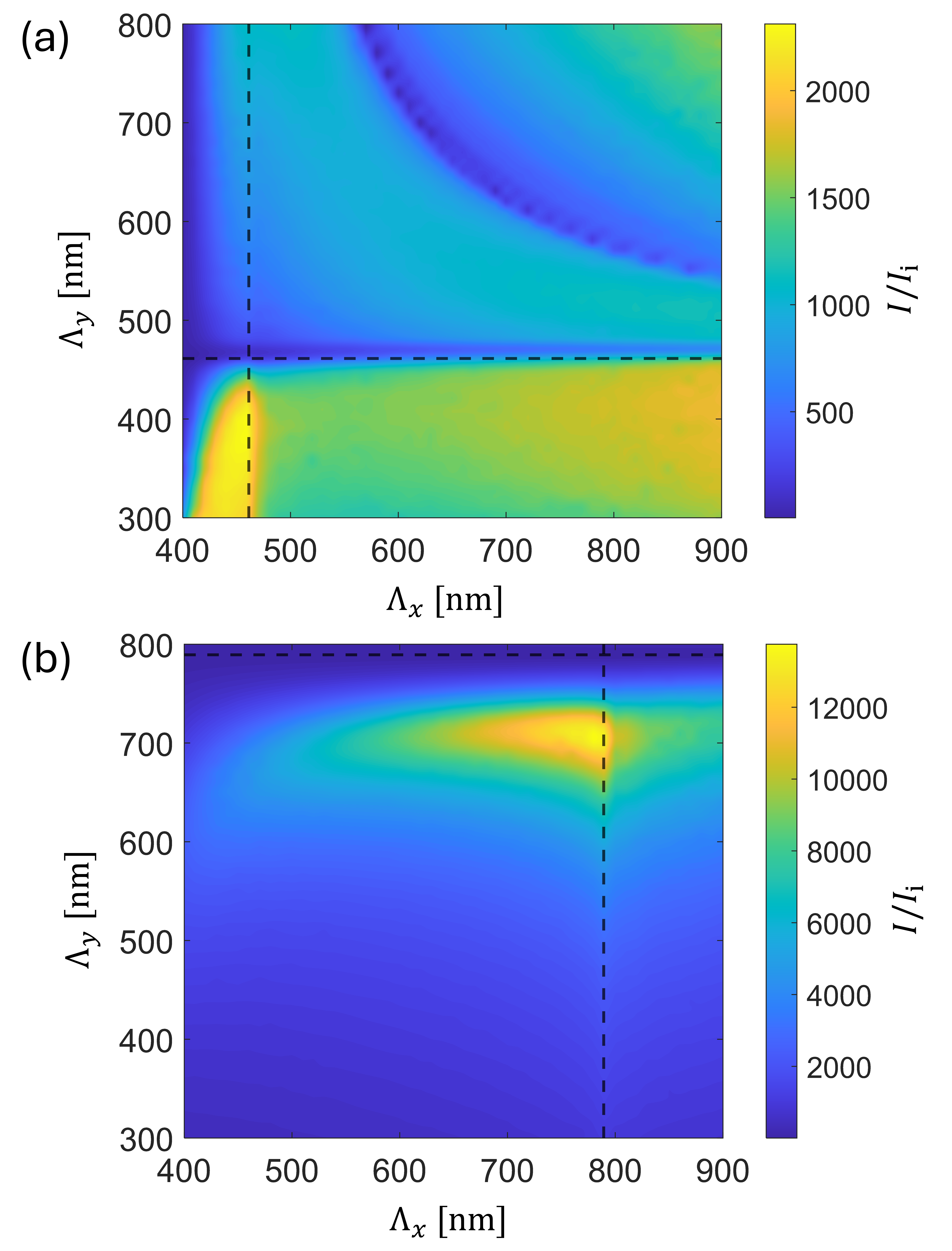}
    \caption{The near-field intensity enhancement in a gold-dimer array as a function of periods $\Lambda_x$ and $\Lambda_y$ at (a) $\lambda =$ 692 nm and (b) $\lambda  =$ 1184 nm.}
    \label{2D}
\end{figure}
\par In Fig. \ref{2D}, we show the gap enhancement calculated for a dimer array with varying periods $\Lambda_x$ and $\Lambda_y$ at the vacuum wavelengths of the octupole and dipole resonances of 692 and 1184 nm, respectively. The corresponding wavelengths in glass are 461 and 789 nm. The lattice resonances take place when the lattice period is close to these values. Hence, in Fig. \ref{2D}, these values are shown by the vertical and horizontal dashed lines. Obviously, close to the intersection of these two lines, two guided-mode lattice resonances (corresponding to two waves propagating along the $x$- and $y$-axis) can hybridize with the plasmon resonance and contribute to the near-field enhancement by the particles. The hybridization leads to Fano-like profiles of field enhancement  along the $x$- and $y$-directions. From Fig. \ref{2D} (a), we see that the hybridized lattice-plasmon resonance of the octupole nature yields a highest near-field enhancement at parameters $\Lambda_x$ and $\Lambda_y$ close to their previously optimized values of 430 nm and 300 nm, respectively. A similar behavior is observed for the dipole resonance at 1184 nm wavelength. The enhancement is maximized approximately at $\Lambda_x =$ 780 and $\Lambda_y =$ 700 nm. The dipole resonance can be used if the near-field enhancement of infrared light is of interest. Note that here, the maximum enhancement factor is larger than in the case of octupole excitation characterized in Fig. \ref{2D}(a). However, the periods $\Lambda_x$ and $\Lambda_y$ have increased approximately by a factor of 2 each, which means that the density of the hotspots has decreased by a factor of $ca.$ 4, so that the surface-averaged enhancement factor at its maximum is expected to be on the same order of magnitude at 692 nm and 1184 nm wavelengths.
\section{Conclusion}
In this work, we have shown that the array factor plays a significant role in the near-field enhancement by plasmonic particles, with a clear distinction between multipole resonances of different orders excited in the particles. As an example, the dipole polarizability vanishes when the period decreases to far subwavelength values, while the octupole polarizability rapidly grows. This effect was obtained analytically, considering an array of point scatterers. It is explained by efficient excitation of octupole moments by inhomogeneous near-fields of the neighboring particles. It can find applications in the fields of surface-enhanced fluorescence and Raman scattering, as well as in optical sensing and nonlinear nano-optics. We have presented a design of gold dimers providing particularly high near-field enhancement at the octupole resonance in the visible spectral range when arranged in a two-dimensional periodic array. We have proposed a method to numerically evaluate the relevant polarizabilities for arrayed particles and achieved reasonable agreement between the numerical and analytical calculations of their values at different periods of the array.
\par In this study, we used a disc dimer geometry to achieve a high near-field enhancement in the visible range through octupole plasmon resonances in a rectangular array of the particles. However, many other shapes of the particles and the unit cells of the arrays, as well as other higher-order multipoles, can be studied in the future. Furthermore, to provide a better confinement of the scattered fields at the array's surface, one can place the particles on a thin slab waveguide. Other strategies to further increase the enhancement factor by higher-order multipole resonances can also be proposed and verified in the future.  In general, higher-order plasmon resonances in arrayed metal particles have not been studied much, and we believe that the field is still open to many scientific and technological discoveries.
\begin{acknowledgments}
This work is part of the Research Council of Finland Flagship Programme, Photonics Research and Innovation [Grant No. 320167 (PREIN Flagship -- Aalto University)]. K.L. was supported by the University of Cologne through the Institutional Strategy of the University of Cologne within the German Excellence Initiative (QM$^2$). The support of the Aalto University School of Science Visiting Professor programme for K.L. is acknowledged.
\end{acknowledgments}
\section*{Data Availability}
The data that support the findings of this article are not publicly available. The data are available from the authors upon reasonable request.
\renewcommand{\theequation}{A\arabic{equation}}
\setcounter{equation}{0}
\appendix*
\section{Derivation of dipole field distribution function $\zeta_d$}
\begin{figure}[t]
    \centering
    \includegraphics[width=\linewidth]{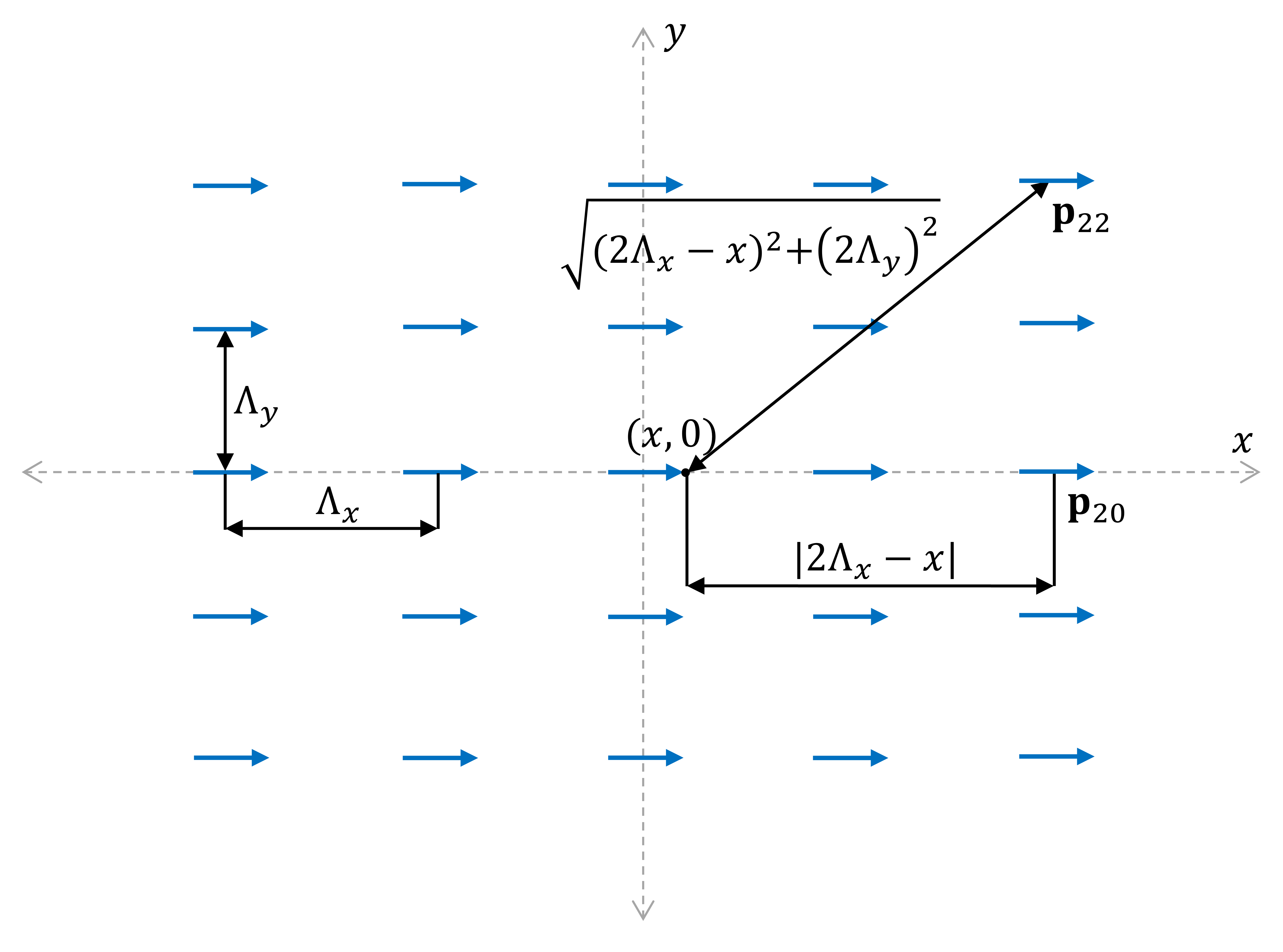}
    \caption{Periodic array of electric dipoles oscillating in phase with each other distributed in the $xy$-plane. The distances from dipoles $\textbf{p}_{20}$ and $\textbf{p}_{22}$ to the observation point $(x, 0)$ are shown by double arrows. In $\textbf{p}_{nm}$, the subindices show the location of the dipole in the array.}
    \label{app}
\end{figure}
Consider an oscillating electric dipole located at the origin of a spherical coordinate system and oriented along the direction $\theta =$ 0. The electric field radiated by the dipole has only the radial and polar vector components given by \cite{novotny2012principles}
\begin{equation}
    E_r=\frac{|p| \space \text{cos}\theta}{4 \pi \epsilon \epsilon_0}\frac{e^{ikr}}{r}k^2\bigg[\frac{2}{k^2r^2}-\frac{2i}{kr}\bigg],
\end{equation}
\begin{equation}
    E_\theta=\frac{|p|\space \text{sin}\theta}{4\pi\epsilon \epsilon_0}\frac{e^{ikr}}{r}k^2\bigg[\frac{1}{k^2r^2}-\frac{i}{kr}-1\bigg],
\end{equation}
where $p$ is the amplitude of the dipole moment and $k$ is the wavenumber in the surrounding medium.
\par Consider next a periodic array of such dipoles distributed in the $xy$-plane, and oscillating in phase (see Fig. \ref{app}). Let the size of the array be $N\times M$. We want to find the total field at a coordinate $(x,0)$ radiated by all the dipoles except the one at the origin of the coordinate system. The fields of the dipoles located on the $x$-axis are purely radial at this point. Hence, using Eq. (A1) and summing these fields gives 
\begin{equation}
\begin{split}
     E_{y=0}&=p\zeta_\text{d}^{(y=0)}\\
    &=p\! \sum_{n=-\frac{N - 1}{2},\space (n\neq0)}^{\frac{N-1}{2}}\! \bigg(\frac{k^2}{2\pi\epsilon \epsilon_0 |n \Lambda_x-x|}\bigg[\frac{1}{k^2(n\Lambda_x-x)^2}\\
    &-\frac{i}{k|n\Lambda_x-x|}\bigg]e^{ik|n\Lambda_x-x|}\bigg).
\end{split}
\end{equation}
Here, the distance from dipole $\textbf{p}_{n0}$ to the considered point $(x, 0)$ is equal to $|n\Lambda_x - x|$. In Fig. \ref{app}, this distance is shown for dipole $\textbf{p}_{20}$. In $\textbf{p}_{nm}$, the subindices $n \in [-(N-1)/2, (N-1)/2]$ and $m \in [-(M-1)/2, (M-1)/2]$ show the position of the dipole in the array. Numbers $N$ and $M$ are assumed to be odd. For a dipole $\textbf{p}_{nm}$ with $m\neq 0$, the distance to the point $(x, 0)$ is equal to $\sqrt{(n\Lambda_x - x)^2 + (m\Lambda_y)^2}$. This distance is in Fig. \ref{app} shown for a dipole $\textbf{p}_{22}$. In addition, both the radial and polar vector components of the electric field are present at that point. However, by symmetry, when the fields of dipoles with positive and negative values of $m$ are summed, only the $x$-component remains. Hence we can sum $x$-components of the fields of the dipoles with $m > 0$ only and double the result. The radial and polar components projected onto the $x$-axis are $E_r \text{cos}\theta$ and $E_\theta \text{sin}\theta$, where $\text{cos}\theta = |n\Lambda_x - x|/\sqrt{(n\Lambda_x - x)^2 + (m\Lambda_y)^2}$ and $\text{sin}\theta = m\Lambda_y/\sqrt{(n\Lambda_x - x)^2 + (m\Lambda_y)^2}$. Using Eqs. (A1) and (A2) and summing the contributions of all the dipoles with $m\neq 0$ to the field at the point $(x, 0)$, we obtain
\onecolumngrid
\begin{widetext}
    \begin{equation}
\begin{split}
     E_{y\neq 0} &= p\zeta^{(y\neq 0)}_\text{d} =p\sum_{n=-\frac{N-1}{2}}^{\frac{N-1}{2}}\bigg(\sum_{m=1}^{\frac{M-1}{2}}\bigg(\frac{k^3}{2 \pi \epsilon \epsilon_0}\bigg[\frac{2(n\Lambda_x-x)^2+(m\Lambda_y)^2}{k^3\big((n\Lambda_x-x)^2+(m\Lambda_y)^2\big)^{3/2}}-\frac{i\big(2(n\Lambda_x-x)^2+(m\Lambda_y)^2\big)}{k^2\big((n\Lambda_x-x)^2+(m\Lambda_y)^2\big)}\\
     &-\frac{(m\Lambda_y)^2}{k\sqrt{(n\Lambda_x-x)^2+(m\Lambda_y)^2}}\bigg]\frac{1}{(n\Lambda_x-x)^2+(m\Lambda_y)^2}e^{ik\sqrt{(n\Lambda_x-x)^2+(m\Lambda_y)^2}}\bigg)\bigg).
\end{split}
\end{equation}
\end{widetext}
\twocolumngrid
\noindent The overall field at the point $(x, 0)$ is $E = E_{y = 0} + E_{y\neq 0}=p\zeta_\text{d}$, from which we obtain the overall function $\zeta_\text{d}$ given by Eq. (19).

\bibliography{Bibliography}
\onecolumngrid
\renewcommand{\theequation}{A\arabic{equation}}

\end{document}